# Bio-Mechanical Model of the Brain for a Per-Operative Image-Guided Neuronavigator Compensating for "Brain-Shift" Deformations

M. BUCKI†, C. LOBOS†‡, Y. PAYAN†

†TIMC Laboratory, UMR CNRS 5525, University J. Fourier, 38706 La Tronche, France
‡Departamento de Ciencias de la Computación, Av. Blanco Encalada 2120, Santiago de Chile, Chile

Corresponding author. Email: Marek.Bucki@imag.fr

## 1 Introduction

Accurate localization of the target is essential to reduce the morbidity during a brain tumor removal intervention. Image-guided neurosurgery is facing an important issue for large skull openings, with intraoperative changes that remain largely unsolved. In that case, deformations of the brain tissues occur in the course of surgery because of physical and physiological phenomena. As a consequence of this brain-shift, the preoperatively acquired images no longer correspond to reality; the preoperative based neuronavigation is therefore strongly compromised by intraoperative brain deformations. Some studies have tried to measure this intra-operative brain-shift. Hastreiter et al. [1] observed a great variability of the brain-shift ranging up to 24 mm for cortical displacement and exceeding 3 mm for the deep tumor margin; the authors claim for a non-correlation of the brain surface and the deeper structures. Nabavi et al. [2] state that the continuous dynamic brain-shift processes evolves differently in distinct brain regions, with a surface shift that occurs throughout surgery (and that the authors attribute to gravity) and with a subsurface shift that mainly occurs during resection (that the authors attribute to the collapse of the resection cavity and to the intra-parenchymal changes). In order to face this problem, authors have proposed to add to actual image guided neurosurgical systems a module to compensate brain deformations by updating the preoperative images and planning according to intraoperative brain shape changes. The first algorithms developed proposed to deform the preoperatively acquired images using image-based models. Different non-rigid registration methods were therefore provided to match intraoperative images (mainly MRI exams) with preoperative ones [3], [4], [5]. More recently, biomechanical models of the brain tissues were proposed to constrain the image registration: the models are used to infer a volumetric deformation field from correspondences between contours [6], [7] and/or surfaces [8] in the images to register. Arguing against the exorbitant cost of the intraoperative MRI imaging devices, some authors have proposed to couple the biomechanical model of the brain with low-cost readily available intraoperative data [9] such as laser-range scanner systems [10], [11] or intraoperative ultrasound [12]. This proposal seems appealing from a very practical point of view, compared with the high cost intraoperative MRI device. However, it gives to the biomechanical model a crucial and very central position. This means that a strong modelling effort has to be carried out during the design of the brain biomechanical model as well as its validation through clinical data. In this article we present the framework we have chosen to tackle the brain-shift problem.

## 2 Materials and methods

Miller has shown [13] that brain tissue has a visco-elastic, non-linear behavior which yields equations difficult to integrate in real-time. A simpler model is thus necessary, especially if tissue resection needs to be modelled. We thus chose a continuous mechanical linear and small deformations model. This hypothesis allows us to model in an interactive way, surgical interactions such as cyst drainage and tissue resection. The linear mechanics PDE solution leads to a linear system KU=F where K is the stiffness matrix, U the displacements at the nodes and F the forces applied on nodes. The general solution of this finite element system can be decomposed as a linear combination of elemental solutions computed for each, specifically labelled, 'pilot node' elemental displacement. The KU=F system is solved using the $LL^t$ decomposition of the sparse symmetric positive-definite matrix K. Before surgery, a patient specific mesh is generated. A set of "pilot nodes" are positioned on anatomical structures easy to identify within pre-operative MRI data. During the intervention, once the pilot nodes positions have been recovered within the echographic images, the global tissue deformation can be computed in real-time and the positions of the structures of interest are corrected intra-operatively as shown in figure 1. Cyst drainage and tissue resection are reflected on the stiffness matrix by eliminating from K the contributions of the elements modelling the cyst or the tumor. The L matrix is then recomputed and the elemental displacement equations are solved in an efficient way using optimized data structures that take advantage of the sparsity of the stiffness matrix K. The nodes of the mesh located in regions where resections are likely to occur are associated to lines and columns in K with greater index. Any modification of such cell within K has a very limited impact on the computation of its triangular factor L.

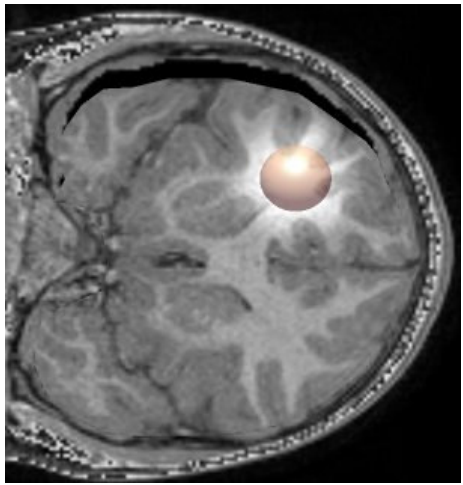

Fig. 1. Effect of gravity on the pre-operative image volume and segmented tumor. The slice shown here simulates an intra-operative MRI.

## 3 Results and discussion

Once all the pre-computations are performed, the "assembly" of the global solution for a 3375 hexahedral mesh of 4096 nodes takes about 20 milliseconds on a 3.0GHz Intel P4. The resection update of the system takes about 5 to 10 seconds on the aforementioned mesh. This approach complies with intra-operative computation time constraints as well as a low-cost or readily-available hardware preference.

## 4 Conclusion

From the biomechanical point of view our linear model is going to be confronted to clinical data. The need to take into account large deformations has to be studied. However, the approach already allows us to model tissue resection in an interactive way.